\documentclass[12pt]{article}
\usepackage{amsmath}
\usepackage{graphicx, psfrag,epsf}
\usepackage{enumerate}
\usepackage{natbib}
\usepackage{url} 

\usepackage{multirow}%
\usepackage{amssymb,amsfonts}%
\usepackage{amsthm}%
\usepackage{mathrsfs}%
\usepackage[title]{appendix}%
\usepackage{xcolor}%
\usepackage{textcomp}%
\usepackage{manyfoot}%
\usepackage{booktabs}%
\usepackage{algorithm}%
\usepackage{algorithmicx}%
\usepackage{algpseudocode}%
\usepackage{listings}%

\usepackage{diagbox}

\usepackage{tikz}
\usepackage{float}
\usepackage[normalem]{ulem}
\usepackage{hyperref}
\usepackage[caption=false]{subfig}

\usepackage{adjustbox}

\definecolor{darkgreen}{rgb}{0.0, 0.5, 0.0}

\newcommand{\blind}{0}

\begin{document}

\def\spacingset#1{\renewcommand{\baselinestretch}%
{#1}\small\normalsize} \spacingset{1}


\if0\blind
{
  \title{\bf Partially Regularized Ordinal Regression to Adjust Teams' Scoring for Strength of Schedule and Complementary Unit Performance in American Football}
  \author{Andrey Skripnikov, Sujit Sivadanam
    \hspace{.2cm}\\
    Natural Sciences Division, New College of Florida\\}
    \date{}
  \maketitle
  }
 \fi

\if1\blind
{
  \bigskip
  \bigskip
  \bigskip
  \begin{center}
    {\LARGE\bf 
    American Football Scores: Using Partially Regularized Ordinal Regression to Adjust for Strength of Opponents, Within-Team Complementary Unit Performance
    
    }
\end{center}
  \medskip
} \fi

\bigskip
\begin{abstract}
American football is unique in that offensive and defensive units typically consist of separate players who don't share the field simultaneously, which tempts one to evaluate them independently. However, a team's offensive and defensive performances often complement each other. For instance, turnovers forced by the defense can create easier scoring opportunities for the offense. Using drive-by-drive data from 2014–2020 Division-I college football (Football Bowl Subdivision, FBS) and 2009–2017 National Football League (NFL) seasons, we identify complementary football features that impact scoring the most. We employ regularized ordinal regression with an elastic penalty, enabling variable selection and partially relaxing the proportional odds assumption. Moreover, given the importance of accounting for strength of the opposition, we incorporate unpenalized components to ensure full adjustment for strength of schedule. For residual diagnostics of our ordinal regression models we apply the surrogate approach, creatively extending its use to non-proportional odds models. We then adjust each team’s offensive (defensive) performance to project it onto a league-average complementary unit, showcasing the effects of these adjustments on team scoring. Lastly, we evaluate the out-of-sample prediction performance of our selected model, highlighting improvements gained from incorporating complementary football features alongside strength-of-schedule adjustments.
\end{abstract}

\noindent%
\vfill

\newpage
\spacingset{1.45} 


\section{Introduction}
\label{sec:Introduction}

In the rapidly growing field of sports analytics, immense progress has been made in developing metrics for performance evaluation, whether for an individual player or a team as a whole. These metrics range from sophisticated formulas leveraging basic box scores to calculate player quality ratings (e.g., offensive/defensive rating in basketball \cite{oliver2004basketball} or quarterback rating in football \citep{oliver2011guide}) to increasingly detailed record-keeping for statistics such as actions or dribbles per touch and the time it takes a quarterback to throw. Additionally, technological advancements have enabled innovations like spatial data tracking \citep{shea2014basketball}. According to \cite{shea2014basketball}, evaluating a player or team involves two primary dimensions: production and context. Most existing findings primarily address production, focusing on the volume of statistical output a player or team generates and the efficiency with which it is achieved. Context, on the other hand, accounts for factors surrounding production, such as the quality of the opponent, the influence of teammates, and game location (e.g., home or away). In this work, we focus on incorporating context into performance evaluation in the sport of American football, where fully contextualized team assessments are particularly critical for reasons outlined below.

The Football Bowl Subdivision (FBS) of Division I American college football, as of 2022, includes 131 participating programs divided into 11 conferences, which are further split into two divisions. Each team plays only 12 opponents during the regular season, primarily within its own division and conference. At season's end, FBS teams are ranked to determine playoff and bowl game participants. However, this scheduling format creates gaps in understanding the relative quality of opposition faced by each team. A similar issue exists in the National Football League (NFL), albeit on a smaller scale. The NFL consists of 32 teams divided into two conferences, each with four divisions. Teams play 16 regular-season games (recently increased to 17), mostly against opponents within their conference and division. Unlike college football, the NFL has clearer criteria on which teams make the playoffs, but the fact that a team doesn't get to face every other team in the league at least once during regular season still leaves an incomplete picture of relative team strengths that year. This aspect sets American football apart from most major sports and leagues, e.g., National Basketball Association, National Hockey Association, Major League Baseball, European soccer leagues, where teams typically play each other at least twice (home and away) during the regular season. Consequently, accounting for strength of schedule is especially crucial when evaluating team performance in both college and professional American football.

To address potential differences in strength of schedule, an ``offense-defense'' model was introduced in \citep{harville1977use}, and later also applied by \citep{govan2009offense} in the contexts of college basketball and professional American football. This model assumes that a team's scoring is a function of its offensive performance and its opponent's defensive performance, thereby adjusting for the strength of the opposition. Additionally, recognizing the significance of home-field advantage \citep{edwards1979home, vergin1999no}, the model in \citep{harville1977use} incorporates whether a team played most of its games at home or away, ensuring further refinement in evaluating team performance.

In this work, beyond adjusting for strength of schedule and home-field advantage when evaluating each team's offense and defense, we also aim to incorporate an adjustment for the complementary nature of the game. Unlike most other sports, American football features offense and defense played by largely non-overlapping groups of players who do not take the field simultaneously. This might suggest that a team’s offensive performance could be evaluated independently of its defensive performance. However, there is reason to believe that the two sides complement each other. For instance, turnovers forced by a team’s defense can create easier scoring opportunities for its offense, while an offense's ability to control the clock can help the defense rest and recover. This concept, referred to as ``complementary football'',is a central focus of this work. One of our primary goals is to quantify and develop an adjustment for this complementary aspect of the game.

We leverage sequential drive-by-drive data from the $2014$–$2020$ FBS college football seasons and the $2009$–$2017$ NFL seasons, where a ``drive''represents a continuous possession of the ball by one team. To identify the most critical features of complementary football in the collegiate game, we employ a tailored regularized estimation technique for ordinal regression with a partial elastic net penalty. Specifically, we utilize the optimization framework introduced in \citep{wurmOrdinalNet} to relax the proportional odds assumption of ordinal regression models. We further extend its penalization scheme, enabling the guaranteed inclusion of proportional-odds coefficients for selected predictors without requiring the inclusion of their non-proportional counterparts. Given the high number of parameters needed to estimate team-level effects that are used to model the strength-of-schedule aspects, this extension makes the estimation task much more feasible. Additionally, since residual diagnostics for categorical models are less developed compared to classical regression, we adopt the surrogate approach introduced in \citep{liu2018residuals} for ordinal regression models. We proceed to enhance the functionality of $R$ package {\em sure} \citep{greenwell2018residuals}, which only applies to proportional odds models, by enabling it to generate residual reasonable diagnostics for non-proportional odds models as well.

\if0\blind
{
Previous work on adjusting for the complementary nature of American football includes \citep{skripnikov2023partially}, who addressed a similar task but relied on cumulative game totals rather than sequential data. This approach failed to capture the direct impact of one unit's performance on the complementary unit's outcome in the subsequent drive. \citep{yurko2019nflwar} developed a framework to estimate the expected points generated by each play based on game context (e.g., down, distance, time remaining), enabling credit to be distributed among a football team’s units. However, their work did not explicitly project performance onto a league-average complementary unit or perform variable selection across statistical categories from the preceding drive while accounting for opponent strength. Related ideas have been explored in other sports. For instance, \citep{kempton2016expected} modeled expected points in rugby plays based on the starting field position and the previous possession's outcome. \citep{boshnakov2017bivariate} analyzed the relationship between goals scored and allowed by a team in soccer matches. However, unlike American football, these sports feature a more continuous flow between offense and defense, with the same players participating in both aspects of the game simultaneously. Consequently, adjusting a team's offensive evaluation based on their defensive quality—and vice versa—is less practical for such sports compared to American football, where offense and defense are played by separate units.
} \fi

\if1\blind
{
Previous work on adjusting for the complementary nature of American football includes [X], who addressed a similar task but relied on cumulative game totals rather than sequential data. This approach failed to capture the direct impact of one unit's performance on the complementary unit's outcome in the subsequent drive. \citep{yurko2019nflwar} developed a framework to estimate the expected points generated by each play based on game context (e.g., down, distance, time remaining), enabling credit to be distributed among a football team’s units. However, their work did not explicitly project performance onto a league-average complementary unit or perform variable selection across statistical categories from the preceding drive while accounting for opponent strength. Related ideas have been explored in other sports. For instance, \citep{kempton2016expected} modeled expected points in rugby plays based on the starting field position and the previous possession's outcome. \citep{boshnakov2017bivariate} analyzed the relationship between goals scored and allowed by a team in soccer matches. However, unlike American football, these sports feature a more continuous flow between offense and defense, with the same players participating in both aspects of the game simultaneously. Consequently, adjusting a team's offensive evaluation based on their defensive quality—and vice versa—is less practical for such sports compared to American football, where offense and defense are played by separate units.
} \fi

The structure of the paper is as follows. Section \ref{sec:MaterialsMethods} details the data and outlines all modeling considerations, including preliminary diagnostics for classic ordinal regression fits, the main partial regularization criteria with relaxation of the proportional odds assumption, residual diagnostics for the selected reduced models, the mechanism of projecting performance onto league-average opponent, homefield factor and complementary unit, wrapping up with the description of methods for out-of-sample predictive performance evaluation of our selected model. Section \ref{sec:Results} presents the variable selection stability analysis, explores the effects of complementary football on scoring outcome probabilities and expected points per drive, showcases the largest shifts in points scored per drive due to various aspects of the adjustment (in particular, the strength of schedule and complementary football), concluding with the comparison of the out-of-sample predictive performance of several models under consideration. Finally, Section \ref{sec:Discussion} provides conclusions and a broader discussion of the main findings, along with potential avenues for future research.
\section{Materials and methods}
\label{sec:MaterialsMethods}

\subsection{Data collection and cleaning}
\label{sec:DataCollectionCleaning}

Play-by-play data for the 2014–2020 FBS seasons were obtained using the {\em cfbfastR} \citep{cfbfastR} package in the $R$ Statistical Software \citep{RCoreTeam}. While the dataset is extensive, significant data cleaning was required to address various inconsistencies and prepare the data for analysis. These issues included lack of continuity in field positions between consecutive plays, mismatches between play yardage values and field position changes, and occasional discrepancies between play descriptions and numerical entries. First, overtime plays were removed due to the distinct format of college football overtimes \citep{wilson2020college}, which essentially eliminates the complementary aspects of the game. Next, despite efforts to correct verifiable inconsistencies by cross-checking play text, yardages, and field positions, many observations remained unverifiable and were subsequently excluded. Given the sequential nature of drives in football, removing a single play or drive could disrupt the overall structure. To preserve the integrity of the data, we opted to drop the entire half of a game containing faulty observations. Since each half begins with a kickoff, independent of prior drives, it can be treated as a continuous, standalone period for measuring the complementary nature of the game. After filtering, $65\%$ of the initial halves were retained, amounting to over $15,000$ drives per season (except for the shortened 2020 season, which included $9,000$ drives). On average, $55$–$60$ drives were available for evaluating each team’s offensive and defensive units. The proportion of problematic data was consistent across seasons ($32\%$–$35\%$), with no particular conference contributing disproportionately to the errors.

The NFL play-by-play data for the 2009–2017 seasons was sourced from Kaggle \citep{Kaggle_NFL_PlayByPlay_Data}. Unlike college football, overtime dynamics in the NFL closely resemble regulation play, so these observations were retained. While the dataset encountered issues similar in nature to those in the college dataset, the proportion of problematic games was significantly smaller. Nearly $98\%$ of all games were retained, with at least $95\%$ retained in any given season. This higher data retention rate allows for a more robust analysis of NFL data. Each NFL team had at the very least 12 games of data available per season, compared to the college dataset, where some teams were left with as few as 4–5 games after excluding faulty observations. As a result, while results will be reported and compared for both leagues, greater emphasis is placed on the NFL due to its larger and more reliable sample size.

The play-by-play data were summarized and grouped by drives, calculating various metrics such as total yards gained, completed and incomplete passes, positive and negative rushing plays, sacks, fumbles, first downs gained, and third-down conversions, among others. To account for such plays as field goals, punts, and kickoffs (collectively known as ``special teams'' in American football) within our offense-defense framework, field goal kicking and kick/punt returning were treated as part of the offense, while their counterparts, such as field goal blocking and kick/punt coverage, were considered part of the defense. 
For a complete glossary of calculated statistics, refer to the Table \ref{tab:GlossaryCompFeatures} in the Appendix.
Since only one unit (offense or defense) from a team is on the field during a single drive, and these units typically alternate between consecutive drives, the drive-by-drive format is particularly well-suited for examining the direct impact of a defense's (or offense's) performance on its complementary unit’s output during the subsequent drive.

Lastly, since some scheduled games for each FBS team in college football traditionally occur against non-major teams from lower divisions, we created a single ``Non-Major'' category to group all such opponents. This approach allows these games to be incorporated into the adjustment mechanism and ranking calculations while avoiding the need to estimate separate parameters for each non-major team, which would otherwise rely on a small and unreliable sample of games.

\subsection{Categorical ordinal nature of points scored during a drive}
\label{sec:CategOrdinalNature}

In American football, a single drive can result in one of five ordinal scoring outcomes, as outlined in Table \ref{tab:ScoringOutcomesDrive}. The highest positive value is $7$ for an offensive touchdown, comprising $6$ points for reaching the opponent's endzone, plus an extra point from a successful kick. Occasionally, the offense may attempt a two-point conversion, which is more challenging than the kick but yields $2$ extra points if successful. Thus, an offensive touchdown can result in $6$ points (missed kick or failed two-point attempt), $7$ points (successful extra point kick), or $8$ points (successful two-point attempt). Despite these possibilities, $7$ points is by far the most common outcome of a touchdown. To simplify the scoring designations, we standardize touchdowns to $7$ points for modeling purposes. The same logic applies to defensive touchdowns, with the difference that they are recorded as negative values for the offense.

\begin{table}
\centering
    \caption{Scoring outcomes of a single drive, their point value designations and occurrence rates in the form of percentage of drives when they occur.}
    \begin{center}
    \begin{tabular}{llll}
      Outcome   &  Points & Drive \% (NFL) & Drive \% (College) \\
      \hline
       1. Defensive Touchdown  & -7\footnotemark[1] & 1\% & 1\%\\
       2. Safety & -2 & $<0.5\%$  & $<0.5\%$\\
       3. No Score & \phantom{-}0 & 65\% & 65\%  \\
       4. Field Goal & \phantom{-}3 & 14\% & 8\% \\
       5. Offensive Touchdown & \phantom{-}7\footnotemark[1] & 20\% & 26\% \\
    \end{tabular}
     \end{center}
    \footnotemark[1]{Can also be 6 or 8 points, albeit rarely.}
    \label{tab:ScoringOutcomesDrive}
\end{table}

Despite each scoring outcome having a numerical value assigned to it, classic linear regression is not appropriate here due to the categorical nature of the mechanism by which these values are produced. Instead, we leverage ordinal regression approaches, where the focus is on modeling the probabilities of these categorical ordinal outcomes. It presents a much more appropriate modeling framework which would then subsequently allow us to designate the numerical values to the categorical outcomes in a weighted fashion, still lending itself to an intuitive interpretation and helping generate college football rankings for expected points per drive scored and allowed. See the supplement for the residual diagnostics of linear regression compared to ordinal regression for our data, demonstrating that the latter is the a much natural fit.

\subsection{Baseline ordinal regression model}
\label{sec:Notation}

Below we introduce the notation, main modeling considerations, and the baseline model which we used as a starting point for our analyses. 

Let $s$ denote the enumeration of the scoring outcomes listed in Table \ref{tab:ScoringOutcomesDrive}, with $s = 1, 2, \dots, 5$. For a total of $n$ teams in the league, let $Y_{ijkl}$ represent the categorical outcome of points scored by the offense of team $i$ against the defense of team $j$ during the $k^{th}$ drive of the $l^{th}$ game between these two teams in a given season, where $i, j = 1, \dots, n$. We assume that $Y_{ijkl}$ follows a multinomial distribution with probability vector $\pmb{\pi}_{ijkl} = (\pi_{1,ijkl}, \dots, \pi_{5,ijkl})^\intercal$, where $\pi_{s,ijkl} = P(Y_{ijkl} = s)$ for $s = 1, \dots, 5$. To capture the ordinal nature of the response variable, we adopt a cumulative probability approach, modeling the odds of the response category being greater than or equal to $s$, for $s = 2, 3, 4, 5$. Incorporating adjustments for strength of schedule, home-field advantage, game context variables and the performance statistics exhibited by the complementary unit during the {\em prior} drive, our baseline model is specified as follows:
\begin{equation}
\begin{split}
\label{eq:ComplementaryBaselineModel}
& Y_{ijkl} \stackrel{iid}{\sim} Multinomial(1,\pi_{ijk}), \\ 
& \log(\frac{P(Y_{ijkl} \geq s)}{P(Y_{ijkl} < s)}) = \mu_s + \alpha_i + \beta_j + \delta h_{ijl} +  \pmb{\phi}^\intercal \textbf{g}_{ijkl} + \xi I.cmp_{ijkl} + \pmb{\gamma}^\intercal \textbf{x}^*_{ji(k-1)l} \times I.cmp_{ijkl}, \\
\end{split}
\end{equation}
where $\mu_s$ denotes the cumulative odds of the respective scoring category $s$ simultaneously for both the offense and defense of what we will call a ``league-average'' opponent. This symmetry arises because as one team's offense scores points, the opponent's defense allows those points. To model each team's offensive and defensive strength, we introduce coefficients $\pmb{\alpha} = (\alpha_1, \dots, \alpha_n)^{\intercal}$ and $\pmb{\beta} = (\beta_1, \dots, \beta_n)^{\intercal}$. Here, $\alpha_i$ represents the margin of offensive improvement for team $i$ over the league-average opponent, while $\beta_j$ captures the defensive margin of improvement for team $j$. We enforce constraints $\sum_{i=1}^n \alpha_i = \sum_{j=1}^n \beta_j = 0$, which anchor the intercept term to the league-average opponent. The term $h_{ijl}$ denotes the home-field indicator for the $l^{th}$ game between teams $i$ and $j$, taking the value $1$ if team $i$ is the home team and $0$ otherwise. The coefficient $\delta$ quantifies the impact of home-field advantage. 

Additionally, $\textbf{g}_{ijkl}$ is a vector capturing game context variables relevant at the start of the drive. These include factors such as the half indicator (1st or 2nd/OT), time remaining in the half, score differential, and pairwise interactions among these variables. Prior work by \citep{yurko2019nflwar} demonstrated the significance of these variables in affecting expected points added during a drive. We confirmed their statistical significance at the $0.05$ level in the baseline model (Equation~\ref{eq:ComplementaryBaselineModel}) for most NFL and college football seasons under consideration. While a three-way interaction term was also examined, it was not found to be statistically significant for any of the seasons analyzed.

As the primary contribution of this work to the application domain, the vector $\pmb{\gamma} = (\gamma_{1}, \dots, \gamma_{C})^\intercal$ represents the effects of all $C$ complementary statistics $\textbf{x}^*_{ji(k-1)l} = (x^*_{1,ji(k-1)l}, \dots, x^*_{C,ji(k-1)l}) = (x_{1,ji(k-1)l} - \bar{\textbf{x}}_1, \dots, x_{C,ji(k-1)l} - \bar{\textbf{x}}_C)$, where team $j$'s offense accumulated these statistics against team $i$'s defense during the $(k-1)^{th}$ drive (i.e., the drive immediately preceding the $k^{th}$ drive) of the $l^{th}$ game between them. The inclusion of data from the preceding drive enables the evaluation of the sequential complementary impact that a team's offense and defense may exert on one another. By centering the complementary statistics through mean subtraction, the intercept terms $\mu_s$ and the offensive/defensive margins ($\alpha_i$/$\beta_j$) can be interpreted as projections onto the league-average complementary unit performance, in addition to the league-average opponent.

Lastly, model (\ref{eq:ComplementaryBaselineModel}) incorporates a component to address the following caveat: approximately $5$-to-$10\%$ of drives in both the college and NFL datasets are either not preceded by any other drive (e.g., at the start of a half or overtime) or follow a drive by the same team's offense (e.g., after a defensive touchdown, where the same offense that allowed the score must return to the field). While such cases do not contribute to the study of direct complementary football impacts, these observations are necessary for incorporating the points scored during these drives into the final estimates of a team's offensive and defensive capabilities. To handle this, model (\ref{eq:ComplementaryBaselineModel}) includes a $\textbf{x}^*_{jikl} \times I.cmp_{ijkl}$ interaction term. Here, $I.cmp_{ijkl}$ is an indicator variable that takes the value $1$ for drives where complementary football effects can be estimated, and $0$ otherwise. For drives where $I.cmp_{ijkl} = 0$, the complementary football effects are effectively nullified, ensuring that the model accommodates these cases appropriately.

All in all, model (\ref{eq:ComplementaryBaselineModel}) accomplishes several objectives simultaneously. First, it enables the projection of each team’s performance onto the same league-average opponent, balanced home/away scheduling, and league-average complementary units. Second, it accounts for key game context variables at the start of each drive, such as the score differential, the half of play, and the remaining time in the half. Unlike the aforementioned projections for strength of schedule, homefield factor, and complementary features, it is less practical to project the points scored on each drive onto a single hypothetical game scenario (e.g., a ``tied game with two minutes left in the second half''). As such, we keep the game context values unmodified in our projections. However, controlling for those game context variables ensures a more considerate estimation for the effects of the aforementioned variables that we do manipulate for projection purposes (i.e., homefield factor, strength-of-schedule, complementary football features). Collectively, our model enables a more comprehensive and contextually informed evaluation of each team’s performance.

\subsection{Preliminary model diagnostics}
\label{sec:PreliminaryDiagnostics}

To assess the suitability of our ordinal modeling approach, we performed checks for multicollinearity and conducted preliminary residual diagnostics for the baseline model (\ref{eq:ComplementaryBaselineModel}) and several of its extensions. The generalized variance inflation factor (GVIF) \citep{fox1992generalized} values were found to be below $2$, suggesting no evidence of multi-collinearity. For residual analysis, we employed the surrogate approach introduced in \citep{liu2018residuals}, specifically designed for ordinal regression models. This method generates surrogate residuals based on the continuous latent variable interpretation of cumulative link ordinal regression models, enabling more intuitive diagnostics of residual patterns. The approach, implemented in the {\em sure} package \citep{greenwell2018residuals}, facilitates the creation of residuals-vs-fitted and quantile-quantile (QQ) plots with pattern expectations identical to those in classic linear regression with a continuous response. After applying this diagnostic approach to our baseline model (\ref{eq:ComplementaryBaselineModel}) across nine NFL seasons and seven college seasons, the results indicated a good model fit with no notable deviations. For further details on the diagnostics for one NFL season, as well as a comparison to an inferior linear regression fit as discussed in Section \ref{sec:CategOrdinalNature}, see the supplementary materials.

Next, given the drive-by-drive data format, which clearly groups observations by game, it is natural to investigate whether within-game error correlation might violate the independence assumption of our baseline model (\ref{eq:ComplementaryBaselineModel}). To address this, we utilized the {\em clmm} function from the {\em ordinal} package \citep{ordinal} to fit a mixed-effects cumulative link model incorporating a random intercept for the {\em game\_id} variable. After fitting this mixed-effects model, we conducted a likelihood ratio test comparing it to the baseline fixed-effects model. The test revealed no statistical evidence of within-group dependence for any of the nine NFL and seven college seasons analyzed.

Therefore, the baseline model (\ref{eq:ComplementaryBaselineModel}) has proven to be a reasonable starting point in terms of modeling fit and adherence to the most critical assumptions. One remaining assumption to evaluate is that of proportional odds. This assumption posits that the effect of a variable on the cumulative odds of category $s_1$ is proportional to its effect on the cumulative odds of category $s_2$ for all $s_1 \neq s_2$. Consequently, it implies that all non-intercept coefficients do not vary by the category $s$ under consideration. While this assumption simplifies the estimation process significantly, it can be overly restrictive in practice. To address this, Section \ref{sec:VariableSelectionOutline} describes the regularized estimation approach we adopted, which aims to balance variable selection, efficient estimation, and relaxation of the proportional odds assumption.

\subsection{Variable selection methodology outline}
\label{sec:VariableSelectionOutline}

As discussed in Section \ref{sec:PreliminaryDiagnostics}, the baseline model (\ref{eq:ComplementaryBaselineModel}) enforces the proportional odds assumption, which can be quite restrictive. Fully relaxing this assumption, on the other hand, would replace the single coefficient vectors $\pmb{\alpha}^{n\times 1}, \pmb{\beta}^{n\times 1}, \pmb{\gamma}^{C \times 1}$ with four separate sets for each cumulative probability category: $\pmb{\alpha}^{n\times 1}_s, \pmb{\beta}^{n\times 1}_s, \pmb{\gamma}^{C \times 1}_s, \ s=2,\dots,5$. This adjustment would significantly increase the number of parameters to estimate, rising from approximately $80$ to $270$ for the NFL and from roughly $275$ to over $1,100$ for college football, thus incredibly complicating the estimation process. To address this challenge, we implemented a regularization approach that incorporates a partial elastic net penalty structure. This method allows for relaxing the proportional odds assumption while keeping the estimation problem computationally manageable.

First, considering that the offensive and defensive margin coefficients ($\pmb{\alpha}$ and $\pmb{\beta}$) already require a substantial number of parameters to estimate (approximately $64$ for the NFL and $260$ for college football), allowing these coefficients to vary by response category would significantly increase the complexity of the estimation task. Moreover, with team ranking considerations in mind, our approach ensures that each team's offensive and defensive coefficients are included in the final model by leaving them unpenalized. This strategy emphasizes the games that teams actually played, rather than prioritizing out-of-sample predictive accuracy typically achieved through shrinkage. Additionally, it virtually eliminates the possibility of rank ties in the resulting estimates.

Second, for the $C$ complementary football features, which require a considerably smaller number of parameters ($15$ in the baseline model for both the NFL and college), we can reasonably relax the proportional odds assumption without significantly impacting the feasibility of estimation. To achieve this, we will employ the semi-parallel optimization approach introduced in \citep{wurmOrdinalNet} and implemented in the {\em ordinalNet} package:

\begin{equation}
\begin{split}
\label{eq:MainRegularizationTask}
      \min_{\substack{\pmb{\mu}, \delta, \pmb{\phi}, \pmb{\alpha}, \pmb{\beta}  \\ \ \ \pmb{\gamma}, \  \{\pmb{\gamma}_{s}\}_{s=2}^{5}}}
    \ -\frac{1}{N^*} \ell(\pmb{\mu}, \pmb{\alpha}, \pmb{\beta}, \delta, \pmb{\phi}, \xi, \pmb{\gamma}, \pmb{\gamma}_2, \dots, \pmb{\gamma}_5; \textbf{y}, h, \textbf{g}, \textbf{x}) + \lambda ((\alpha ||\pmb{\gamma}||_1 + \frac{1}{2} (1-\alpha ||\pmb{\gamma}||_2^2) \ +  \\ 
    \ \ \ \ \ \ \ \ \ \  + \  (\sum_{s=2}^5 (\alpha ||\pmb{\gamma}_s||_1 + \frac{1}{2} (1-\alpha) ||\pmb{\gamma}_s||_2^2)))  
\end{split}
\end{equation}
where $\ell()$ represents the multinomial log-likelihood function which, for all complementary football features $\textbf{x} = (x_1, \dots,x_C)^\intercal,$, includes both the proportional odds coefficients $\pmb{\gamma}$ and the non-proportional coefficients $\pmb{\gamma}_s$ for cumulative odds of response category $s, s=2,\dots,5$. This formulation is derived from a modified version of our baseline model (\ref{eq:ComplementaryBaselineModel}), where the right-hand side equation linking parameters to predictors incorporates the effect of complementary football features as $\pmb{\gamma}^\intercal \textbf{x}_{ji(k-1)l} + \pmb{\gamma}_s^\intercal \textbf{x}_{ji(k-1)l}$ instead of just $\pmb{\gamma}^\intercal \textbf{x}_{ji(k-1)l}$. This adjustment allows the model to select, for each feature, either a proportional odds structure (by setting all non-proportional coefficients to zero) or to relax the proportional odds assumption by incorporating non-proportional coefficients. However, such parameterization results in a model that is not identifiable, which is precisely where the elastic net penalty \citep{zou2005regularization} in the second row of the optimization criteria (\ref{eq:MainRegularizationTask}) plays a critical role. The penalty is a weighted combination of LASSO and ridge penalties, with $\alpha \in [0,1]$ determining the weight distribution: $\alpha=1$ corresponds to a pure LASSO penalty, $\alpha=0$ corresponds to ridge, and $\alpha \in (0,1)$ represents a mixture of the two. \citep{wurmOrdinalNet} demonstrated that despite the overparameterized formulation, the optimization task (\ref{eq:MainRegularizationTask}) is guaranteed to have a unique optimum for all cases except $\alpha=1$ (pure LASSO). Since the primary focus of this work is feature selection, which is particularly well-facilitated by the LASSO penalty, we chose a high $\alpha$ value of $0.99$ to avoid non-unique optima while maintaining a strong emphasis on variable selection. Note that we also considered a larger grid of $\alpha$ values that aligned with our emphasis on the LASSO component ([0.8, 0.9, 0.95, 0.99, 0.999, 0.9999]), but the variable selection results were shown to be robust regardless of $\alpha$ choice. See supplementary materials for more detail. 

Although the main optimization algorithm for the multinomial log-likelihood semi-parallel elastic penalty criteria, similar to task (\ref{eq:MainRegularizationTask}), is implemented in the {\em ordinalNet} package, it does not allow selecting a subset of predictors for which to include non-proportional coefficients. Instead, it requires either including non-proportional coefficients for all predictors or none of them. As discussed earlier in this section, incorporating non-proportional odds for all variables would lead to an overly complex estimation task. This limitation prompted us to extend their methodology to accommodate the partial enforcement of the proportional odds assumption. Specifically, we enforced proportional odds for features not subjected to selection (leaving them unpenalized), such as the offense/defense margin parameters $\pmb{\alpha}, \pmb{\beta}$, game-context variable coefficients $\pmb{\phi}$, and the home-field factor parameter $\xi$. In contrast, complementary football features—the primary focus of this work—were subjected to a non-proportional odds formulation. Additionally, these features were integrated into a penalization scheme that enabled both the selection of relevant complementary football features and the partial relaxation of the proportional odds assumption.

To select the tuning parameter value $\lambda$, we employed the classic $10$-fold cross-validation (CV) within each NFL and college season. Specifically, we selected the largest $\lambda$ value that achieved test performance within one standard deviation of the best test performance value, hence favoring a sparser model (“sparse CV”). This approach emphasizes the selection of simpler models with the most critical variables, effectively ensuring that the chosen complementary football features would also be selected by any other commonly used criteria. For the test performance metric, we used the out-of-sample multinomial log-likelihood $\ell()$, which was also utilized in the main optimization task formulation (\ref{eq:MainRegularizationTask}), with higher values indicating a better fit. To address the randomness inherent in the $10$-fold CV procedure—arising from the random partitioning of the data—we conducted three replicates of $10$-fold cross-validation for each season and reported the cumulative findings across all replicates and seasons. For the final evaluation of selection stability in both the NFL and college datasets, we calculated the proportion of times each complementary football feature (and its proportional/non-proportional odds coefficient) was selected across all CV replicates and seasons, taking into account the coefficient signs as well. Additional details on how the sparse CV estimate was chosen for each $10$-fold CV replicate and season can be found in the supplementary materials.

\subsection{Binarized Surrogate Residual Diagnostics for the Selected Model}
\label{sec:ResidualDiagnostics_SelectedModel}

Having identified the most consistent features of complementary football, we refitted the ordinal regression models to only include those features. This step allows us to study the nature of the selected complementary football effects across the years, guaranteeing consistency in ranking adjustment patterns. Given that we ended up consistently selecting some non-proportional odds coefficients for cumulative odds of just one response category, and that there were no packages implementing such an exact specification, we further enhanced the {\em ordinalNet} package \citep{wurmOrdinalNet} to allow for their algorithm to yield such a fit.

To conduct a post-hoc analysis of the selected model's quality of fit, much like during the preliminary analysis, we referred to the surrogate residuals approach \citep{liu2018residuals} carried out in the {\em sure} package \citep{greenwell2018residuals}, but this time with a methodological modification. It's important to note that the surrogate approach mechanism is capable of yielding a reasonable single set of surrogate residuals for the entire ordinal regression fit solely in the case of a proportional odds approach. Therefore, to make it more appropriate for our scenario of a non-proportional odds model, we calculated surrogate residuals for our model's fit to cumulative odds of each scoring outcome category separately. That meant treating each fitted equation from model (\ref{eq:ComplementaryBaselineModel}) as a separate binary logistic regression, and subsequently producing a separate set of residuals and diagnostic plots for each such equation, which the surrogate approach allowed. We called it a ``binarized diagnostics'' approach, and proceeded to enhance the source code of the {\em sure} package in order to incorporate that functionality.

\subsection{Post-hoc analysis of selected features; Projection mechanism}
\label{sec:PosthocAnalysis}

To study the nature of the effect the selected complementary football features had on scoring, we illustrated the impact of a non-scoring turnover on the expected points, across all seasons of both the NFL and college. In particular, it reported the projected additional points per drive that an offense should expect if the drive was preceded by a non-scoring turnover (at the average post-turnover starting field position) as opposed to not. Additionally, we produced prediction plots that demonstrate the relationship between the post-turnover starting position and the probability of each scoring outcome category.

To calculate the projections of each team's points per possession scored (offense) and allowed (defense) onto the league-average opponent, complementary unit, and home/away game balance, we followed these steps. First, we estimated the probabilities of their offensive and defensive performance in scoring category $s$ for each drive by replacing their actual opponent with the league-average opponent. This was achieved by setting all team-level coefficients to 0 (representing the "average" category, as defined by our contrasts), replacing the complementary feature values with league averages, and setting the home-field factor to the league-average proportion of scheduled home games. The game context variables, as discussed at the end of Section \ref{sec:Notation}, were left unmodified to avoid projecting onto an arbitrary hypothetical game scenario.  To calculate the expected points from the estimated probabilities, we computed the weighted average of point denominations for each scoring outcome category (see Table \ref{tab:ScoringOutcomesDrive}). Finally, to obtain the projection for a team's points scored or allowed per drive across an entire game or season, we averaged these single-drive projections over the respective game or season. 

To calculate the projection based solely on strength of schedule, we followed the same steps, except the complementary feature values were left unchanged to avoid projecting onto the league-average complementary unit. We then reported the difference between the values projected only onto the league-average opponent and home-field factor, and the values additionally projected onto the league-average complementary unit.

Lastly, to quantify the uncertainty around the estimated complementary football effects and projections, we employed bootstrap resampling. For each NFL and college season, we generated 1000 bootstrap resamples of the original data and used the 2.5\% and 97.5\% quantiles of the bootstrap replicates to construct 95\% confidence intervals.  Specifically, we applied the bootstrap method to create 95\% confidence bands for the extra points expected due to a non-scoring turnover occurring on the preceding drive compared to when it did not. This was achieved by taking the quantiles of all such extra point projections for each team across the bootstrap replicates. Additionally, we used the bootstrap to generate 95\% confidence bands for the shifts in points-per-drive values resulting from projecting onto the league-average complementary unit, as opposed to only the league-average opponent and home-field factor. These bands were derived for each team from the quantiles of their respective projection shifts across the bootstrap replicates.

\subsection{Out-of-Sample Performance of the Selected Model}
\label{sec:OutOfSamplePerformance_SelectedModel}

For illustrating the out-of-sample predictive performance improvement gained from our incorporation of the complementary football features, we conducted a 10-fold cross-validation (CV) to compare three ordinal regression models with the following hierarchy of feature sets. First model only included game context variables (score differential, half of play, time left in the half), second model added strength of schedule team-level variables, and third was our final model, which also adds the selected complementary football features. For the performance metric we converted the predicted scoring category probabilities into expected points $\hat{y}$ via weighting them by point value designations from Table \ref{tab:ScoringOutcomesDrive}, and used Mean Absolute Error (MAE) to compare those with the actual points scored during the drive $y$: $MAE = mean(|\hat{y} - y|)$. We plotted the cross-validation MAE values for each season of NFL and college, providing the 10-fold CV standard error bars around the MAE point estimates, which were calculated as: $SE = sd(CV.MAE)/\sqrt{10}$. 

Lastly, besides out-of-sample predictions for expected points, we leveraged a diagnostic in strong analogy with \citep{yurko2019nflwar} to illustrate how well-calibrated the final ordinal regression model was in predicting each respective scoring event (defensive touchdown, safety, no score, field goal, offensive touchdown). The only difference in our methodology was doing the calibration within each season, as opposed to their ``leave-one-season-out'' approach. That was dictated by the fact that, unlike \citep{yurko2019nflwar}, our model utilized team-level factor variables to model the effects of strength of schedule in a particular year, and there's no guarantee that teams truly remain the same year-to-year due to things like roster changes, coach and front office firings, students graduating or declaring for the NFL draft (in college), etc. Hence, it didn't make as much sense to leave an entire season out as a test set, and instead we simply did a classic 10-fold cross validation within each respective season. 
\section{Results}
\label{sec:Results}

In this section, we begin by presenting the results of the optimization task (\ref{eq:MainRegularizationTask}) and evaluating the selection stability of complementary football features. After identifying the most consistently selected features, we conduct the binarized surrogate residual diagnostics to confirm that the selected model constitutes an appropriate fit to the data. Subsequently, we illustrate the exact nature of the impact that the selected complementary features have on scoring. Then we showcase the strongest value shifts in points score/allowed per drive that result from projecting each team's offensive and defensive performance onto the league-average complementary unit, in addition to the strength-of-schedule and home-field adjustments. Lastly, we conclude with an illustration of our model's out-of-sample predictive performance compared to approaches that don't include complementary features, confirming the importance of accounting for the latter.

\subsection{Selection stability analysis}
\label{sec:FeatureSelectionStabilityAnalysis}

\begin{figure*}[t]
\centering
\subfloat[Selection stability\label{fig:NFL_SelectionStability}]
{\includegraphics[scale=0.7]{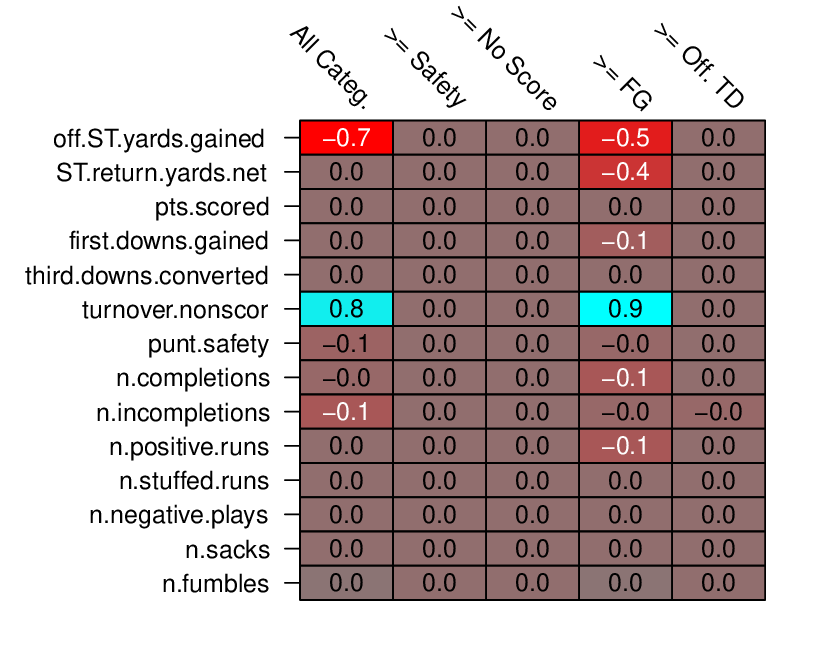}}
\subfloat[Starting field position\label{fig:NFL_StartPos}]{\includegraphics[scale=0.65]{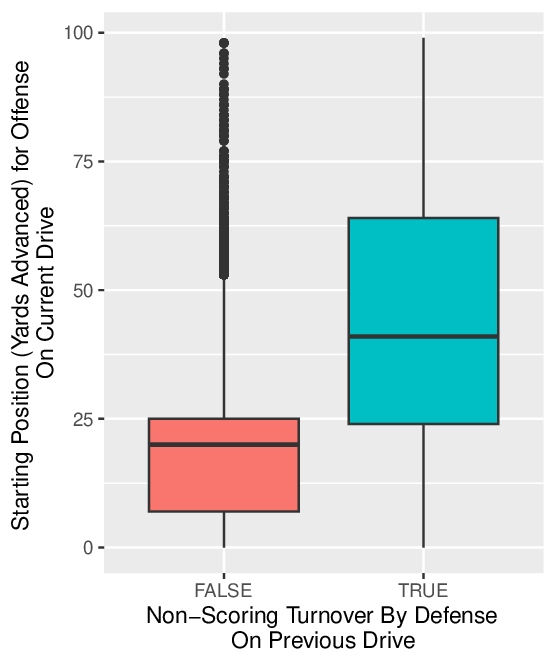}}
\caption{(a) Proportion of cross-validation replicates for the 2009–2017 National League seasons in which each complementary football statistic (rows) was selected based on its impact on scoring during the subsequent drive, including the sign of the coefficient. Each statistic includes a proportional odds coefficient (first column) and non-proportional odds coefficients (columns $2$–$5$, by response category). (b) Box plots of starting field positions (yards advanced toward the opponent’s goal line) for the offense, based on whether their defense forced a non-scoring turnover on the previous drive. For returned punts and kickoffs, the starting field position is measured from the point where the returner caught the ball.
}
\label{fig:NFL_SelectionStability_StartPos}
\end{figure*}

In Figure \ref{fig:NFL_SelectionStability}, we present the selection stability results of the optimization task (\ref{eq:MainRegularizationTask}) for each complementary football statistic (rows) in the case of the 2009–2017 NFL seasons. Specifically, we illustrate the consistency with which the proportional odds coefficient, affecting all scoring categories (first column), and the non-proportional odds coefficients, impacting specific cumulative odds (columns $2$ through $5$), were selected across all cross-validation replicates, including the signs of the respective coefficients. Note that, whenever selected, the respective feature's coefficient had the same sign across all replicates, allowing us to incorporate that sign along with the reported proportions.

Focusing initially on the first column, which represents the effects on all scoring categories, we observe that the non-scoring turnover indicator was selected most consistently (at least $80\%$ of the time). When selected, it always exhibited a positive effect on scoring, indicating that a non-scoring turnover generated by the defense on the previous drive typically leads to more points scored by the offense on the current drive. Meanwhile, yards allowed by the defense (equivalent to yards gained by the opposing offense, {\em off.ST.yards.gained}) were selected in $70\%$ of the cases and consistently showed a negative impact on the complementary offense's scoring. Both of these effects are intuitive when considering their implications for the starting field position of the complementary unit.

Using the example of non-scoring turnovers forced by the defense, Figure \ref{fig:NFL_StartPos} illustrates how these turnovers tend to result in a more advanced starting field position for the offense. The median starting position of 41 yards after a forced turnover is 21 yards better than in its absence, with the entire interquartile range of starting positions being higher than when no such turnover occurs. Regarding yards allowed, defenses that allow more yards tend to put their offenses in less advantageous starting positions, whereas defenses more successful at stopping the opponent's advance create better starting positions for their offenses. Given this reasoning, the positive effect of turnovers and the negative impact of yards allowed on offensive scoring are intuitive. These findings were corroborated using 2014–2020 FBS college football data (see supplementary materials for details), where non-scoring turnovers were selected most consistently and demonstrated a clear improvement in starting position for the offense following a defensive turnover.

Among other consistently selected coefficients, Figure \ref{fig:NFL_SelectionStability} shows that non-scoring turnovers were selected $90\%$ of the time as having a non-proportional effect on the odds of scoring at least a field goal on the next drive (see the ``$\geq FG$'' column). With the non-proportional coefficient being positive, this indicates that turnovers have an even stronger impact when distinguishing drives that ended in an offensive score (field goal or touchdown) from those that did not. This result can be intuitively explained: as shown in Figure \ref{fig:NFL_StartPos}, turnovers often result in a favorable starting field position. In many cases, this starting position is close to the kicker's ``field goal range''—the distance from which they can reasonably attempt a successful kick. While this range varies by kicker, NFL kickers tend to robustly convert field goals starting from about 70–75 yards into the opponent's territory (25–30 yards from the opponent's end zone), whereas in college football, this range is often closer to 75–80 yards. With a median starting position of 41 yards post-turnover in the NFL, the offense needs to gain only an additional 30 yards, or even fewer if the kicker is particularly good, to set up a comfortable field goal attempt.

In contrast, reaching the opponent's end zone for a touchdown requires the offense to gain approximately 60 yards. This task becomes increasingly challenging as the offense approaches the end zone due to the confined space. The end zone is only 10 yards deep, and offensive players cannot step out of bounds, which limits options for deep passing plays and maneuvering. While most offenses tend to score a touchdown more often than not once they are near the opposing end zone, the compressed field makes it far from a guarantee. Conversely, successfully kicking a field goal from such distances is much more reliable. Given this context, it is unsurprising that a good starting position has a particularly strong effect on the likelihood of scoring at least a field goal, but less so on the likelihood of scoring a touchdown, which depends heavily on the offense's ability to execute effectively in the ``red zone'' (within 20 yards of the opponent's end zone). This finding was also confirmed using 2014–2020 college football data.

Given that the aforementioned results clearly highlight the importance of starting position, we aimed to explicitly account for it in our model. Specifically, we wanted to include it for scenarios where the starting position results from the defense turning the ball over {\em directly} to the offense, helping distinguish between cases where a turnover actually leads to a favorable starting position and those where it does not (as shown in Figure \ref{fig:NFL_StartPos}, the starting positions after a turnover can vary considerably). The turnover scenarios we focus on are: non-scoring interceptions, non-scoring lost fumble recoveries, missed field goals, and turnovers on downs. Those are all the scenarios where the offense takes over at the exact spot where the defense left the ball for them. We focus on these cases because, in other scenarios—such as punts and kickoffs—the starting position for the offense is largely determined by the outcome of the punt or kickoff itself. These plays are transitory in nature, where most of the time the possession is expected to change hands between the teams \textit{during} the play rather than on consecutive plays. Although we still acknowledge special teams plays within our offense-defense modeling framework (including metrics such as special teams return net yards, a punt/safety indicator, and a {\em pts.scored} indicator for kickoffs, as these occur after points have been scored), the field position dynamics of these plays are more complex to incorporate in a way that directly credits the complementary unit for creating it. This is especially true when compared to the more straightforward non-scoring turnover scenarios discussed earlier.

To incorporate post-turnover field position, we included a combined {\em turnover.nonscor $\times$ start.pos} term in our initial set of complementary features, where {\em start.pos} represents the offense's starting position following the turnover. Since {\em turnover.nonscor} is an indicator variable ($=1$ if there was a non-scoring turnover on the previous drive, $=0$ otherwise), this term ensures that the starting position only influences the model when a turnover occurs.

As can be seen on Figure \ref{fig:fig_SelectionStability_W_ST_POS_INTERACTION} of the Appendix, after incorporating this term into our main optimization task (\ref{eq:MainRegularizationTask}), it was selected in $100\%$ of all cross-validation replicates across all NFL and college seasons under consideration. Specifically, similar to the non-scoring turnover indicator shown in Figure \ref{fig:NFL_SelectionStability}, both the proportional odds coefficient and the non-proportional odds coefficient for scoring at least a field goal were selected. However, the effect of yards allowed by the defense was not as consistently selected. Therefore, for our final selected model, we chose the non-scoring turnover indicator and the term combining it with the post-turnover starting position, including the non-proportional coefficients for the category of scoring at least a field goal.



\subsection{Binarized Surrogate Residual Diagnostics}

\begin{figure*}[t]
\centering
        \includegraphics[scale=0.8]{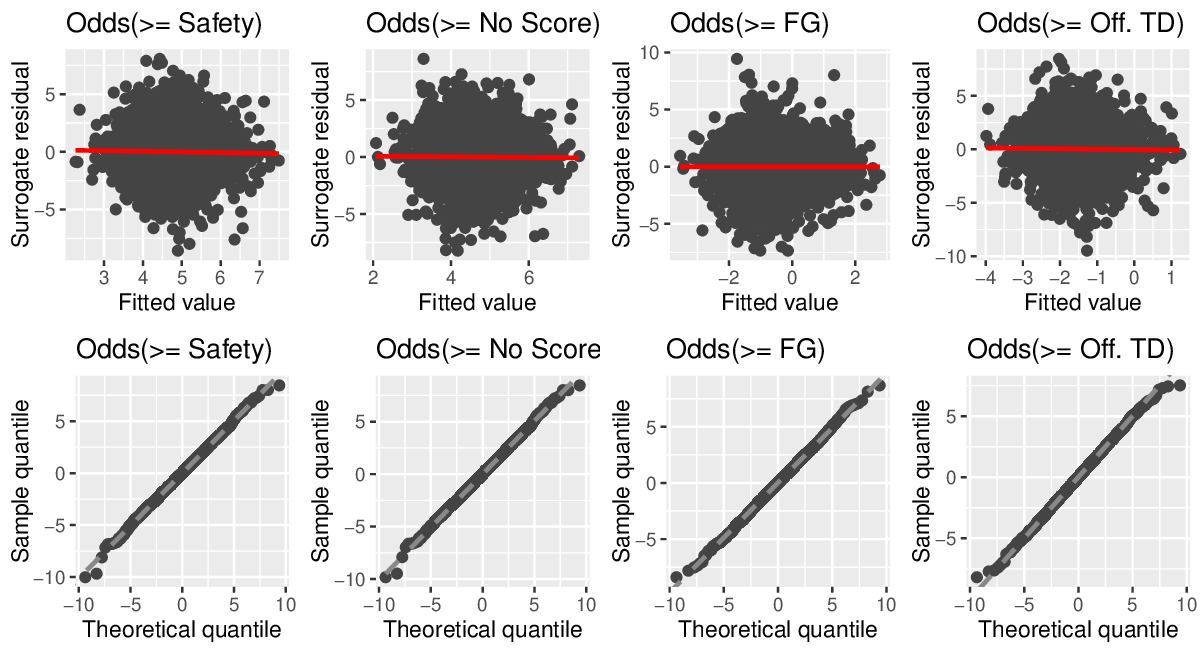}
\caption{Binarized surrogate residuals-vs-fitted plots (top row) and quantile-quantile (QQ) residual plots (bottom row) for the non-proportional cumulative odds model fitted to the 2009 NFL season data, accounting for the most consistent complementary football features. The plots are organized by the cumulative odds being estimated.}
\label{fig:2009NFL_SurrogateResidualPlots}
\end{figure*}

To confirm that our reduced non-proportional odds model properly fits the data, we applied the binarized diagnostics approach, inspired by the surrogate residuals method used in our preliminary analysis \citep{liu2018residuals}, as described in Section \ref{sec:PosthocAnalysis}. As shown in Figure \ref{fig:2009NFL_SurrogateResidualPlots}, both the residuals-vs-fitted and QQ plot patterns indicate a good fit for the 2009 NFL season. As a reminder, the purpose of surrogate residuals is to closely emulate the diagnostics of classic linear regression. Therefore, the patterns we look for to identify a good fit are the same as those for classic regression's residuals-vs-fitted and QQ-plot patterns. Specifically, the residuals-vs-fitted plots show the residual trend line centered at 0, which suggests that the linear effects are appropriate, while the variance of the residuals remains consistent. In the QQ-plots, the sample quantiles align with the theoretical ones for cumulative odds equations across all ordinal response categories. This pattern holds for all NFL and college seasons under consideration (see supplementary materials for an analogous plot of the 2014 FBS college football season).

\subsection{Nature of the Complementary Effects}

\begin{figure*}[h]
\centering
\subfloat[Turnover effect on scoring\label{fig:Extra_Points_After_Turnover_Plot}]
{\includegraphics[scale=0.6]{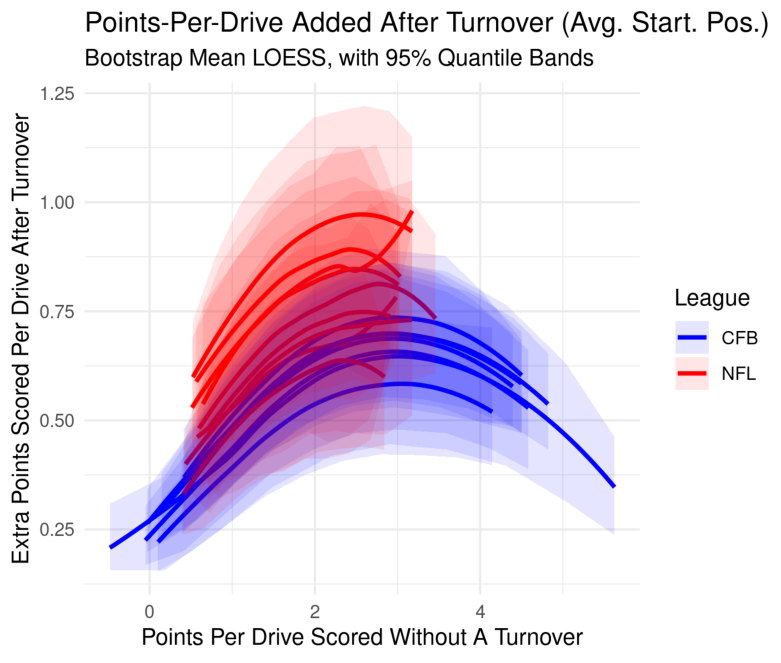}} \ \ \ \
\subfloat[Starting field position effect on scoring \label{fig:NFL_StartPos_Effect_on_Scoring}]{\includegraphics[scale=0.74]{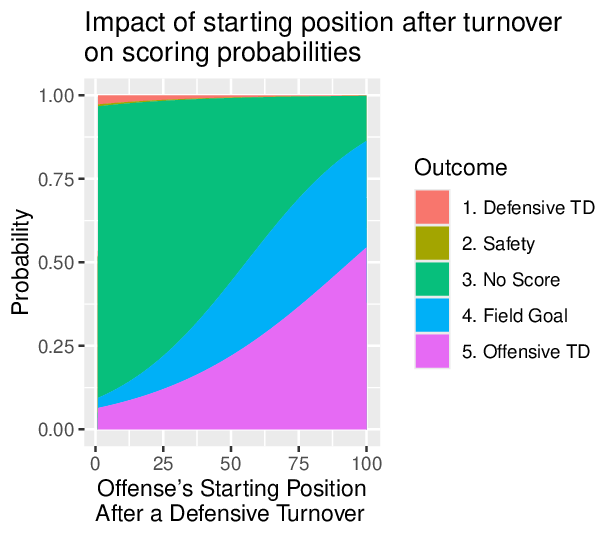}}
\caption{(a): Relationship between the offense's points per drive scored without a non-scoring defensive turnover on the previous drive (the ``baseline'') and the additional points expected after such a turnover (from the average post-turnover starting position) for both college ("CFB") and the NFL, projected onto a league-average opponent. Each smoothed curve represents a season. (b): Effect of post-turnover starting position on the probability of each scoring outcome for the league-average offense facing a league-average defense during the 2009 NFL season.}
\label{fig:Nature_Of_Effects_On_Scoring}
\end{figure*}

The effects of a defensive turnover and post-turnover starting position on offensive scoring are depicted on Figures \ref{fig:Extra_Points_After_Turnover_Plot} and \ref{fig:NFL_StartPos_Effect_on_Scoring}, respectively.

Figure \ref{fig:Extra_Points_After_Turnover_Plot} illustrates the additional points per drive that offenses tend to gain when a drive follows a non-scoring turnover by their defense, compared to drives not preceded by such turnovers (the ``baseline''). For nearly all seasons in both college and the NFL, there is a parabolic relationship: the gain in expected points per drive from a non-scoring takeaway increases up to $0.6$–$1.0$ points. However, once offenses start averaging more than $2$–$2.5$ points per drive as their baseline, the additional points begin to decline, largely due to the upper-bound limit of $7$ points per drive. Interestingly, there are subtle differences between college and professional leagues. In the NFL, offenses are rarely projected to score more than $2.5$–$3$ points per drive without a turnover, whereas some college teams exceed $4$ points per drive in such cases. This likely reflects the stronger parity in the professional league, while the talent distribution across college teams is more uneven. Conversely, NFL offenses appear to capitalize more effectively on turnovers, gaining more additional points than college offenses in the $0.75$–$2.5$ points-per-drive range without a turnover.

Figure \ref{fig:NFL_StartPos_Effect_on_Scoring} highlights the effect of starting position following a non-scoring turnover. There is a steep increase in the probability of scoring at least a field goal (the combined purple and blue areas) as the offense's starting position advances further into the opponent's territory. Unsurprisingly, this corresponds to a decrease in the probabilities of no score or a negative scoring event for the offense (such as a defensive touchdown or safety). This result confirms the intuition that the field position provided by the defense through a turnover plays a pivotal role in the offense's chances of scoring on the subsequent drive.

\subsection{Projection shifts in points scored per drive (offense)}
\label{sec:RankingShifts}

In this section, we showcase some of the largest shifts in projected points per drive resulting from adding an adjustment for complementary football features {\em on top} of the strength-of-schedule and homefield adjustments (for more details on the projection mechanism, refer back to Section \ref{sec:PosthocAnalysis}). Subsequently, we also demonstrate the relative magnitudes of shifts attributable to strength-of-schedule adjustments compared to complementary football adjustments, on both a season-long and a single-game level. It is important to note that the adjustments for each season were performed using only that respective season's data, as it is unreasonable to assume that every team remains the same year-to-year due to changes to player rosters, coaching staff, etc.

Table \ref{tab:SeasonLongShiftsOffenseNFL} lists the five most positive and most negative complementary football adjustment shifts in points scored per drive by NFL offenses across the 2009–2017 seasons. The offense experiencing the strongest positive shift was the Chicago Bears in 2016, with an increase of 0.08 points per drive compared to their strength-of-schedule projection, while other cited offenses experienced similar increases of 0.06–0.07 points. Notably, these offenses were complemented by defenses that ranked near the bottom of the league in post-turnover starting position (none ranked higher than 28th out of 32 that year), with most also performing poorly in generating turnovers (ranking no higher than 31st). The only exception on the latter was New England in 2017, whose defense was above average at generating turnovers that season. However, this highlights that merely generating turnovers may not provide significant value unless they result in favorable starting field position for the offense. New England's defense ranked 31st in post-turnover starting field position, averaging just 36 yards of advancement and leaving their offense with 64 yards to go. To sum up, projecting these teams to a league-average complementary unit (defense, in this case) with respect to both post-turnover starting position and turnover generation tendency resulted in a notable increase in value, given how poorly ranked most of these defenses were in key complementary football metrics.

 \begin{table*}[t]
    \centering
       \caption{The five most positive (top) and five most negative (bottom) shifts in points scored per drive by the offense over a season in the National Football League, along with 95\% bootstrap quantile confidence intervals, corresponding league ranking shifts, and complementary unit performance rankings that season. These shifts are relative to adjustments made only for strength of schedule and homefield factors.}

       \medskip
\begin{adjustbox}{max width=\textwidth}
 \begin{tabular}{rrrcc} 
Team \ & Year \ & Drives \ & Points Scored (Per Drive) & Defense Statistics (Per Drive) \\ \hline 
 \begin{tabular}{r} \\ \\ CHI \\ IND \\ NE \\ HOU \\ DAL  \\ .... \\ KC \\ GB \\ NYG \\ CAR \\ HOU \end{tabular} & 
 \begin{tabular}{r} \\ \\ 2016 \\ 2012 \\ 2017 \\ 2010 \\ 2015  \\ .... \\ 2013 \\ 2009 \\ 2012 \\ 2015 \\ 2011 \end{tabular} & 
 \begin{tabular}{r} \\ \\ 176 \\ 183 \\ 174 \\ 176 \\ 180  \\ .... \\ 209 \\ 164 \\ 183 \\ 201 \\ 152 \end{tabular} &
 \begin{tabular}{rr} \\ Val (Shift, 95\% CI) \ \ & Rk (Shift) \\ \hline 1.46 ($\textcolor{darkgreen}{\Uparrow}$ 0.08, [0.07, 0.11]) & 27 ($\textcolor{darkgreen}{\Uparrow}$ 4) \\ 1.72 ($\textcolor{darkgreen}{\Uparrow}$ 0.07, [0.06,  0.09]) & 16 ($\textcolor{darkgreen}{\Uparrow}$ 2) \\ 2.63 ($\textcolor{darkgreen}{\Uparrow}$ 0.07, [0.05, 0.08]) & 1 ($\textcolor{blue}{=}0$) \\ 2.26 ($\textcolor{darkgreen}{\Uparrow}$ 0.06, [0.05, 0.08]) & 2 ($\textcolor{blue}{=}0$) \\ 1.40 ($\textcolor{darkgreen}{\Uparrow}$ 0.06, [0.04,  0.08]) & 26 ($\textcolor{darkgreen}{\Uparrow}$ 1) \\ ... & ... \\1.45 ($\textcolor{red}{\Downarrow}$ 0.17, [-0.20, -0.15]) & 21 ($\textcolor{red}{\Downarrow}$ 3) \\ 1.90 ($\textcolor{red}{\Downarrow}$ 0.15, [-0.18, -0.13]) & 8 ($\textcolor{red}{\Downarrow}$ 1) \\ 2.03 ($\textcolor{red}{\Downarrow}$ 0.14, [-0.16, -0.12]) & 8 ($\textcolor{red}{\Downarrow}$ 5) \\ 1.89 ($\textcolor{red}{\Downarrow}$ 0.14, [-0.16, -0.12]) & 11 ($\textcolor{red}{\Downarrow}$ 3) \\ 1.85 ($\textcolor{red}{\Downarrow}$ 0.13, [-0.15, -0.10]) & 13 ($\textcolor{red}{\Downarrow}$ 6) \\  \end{tabular} & 
 \begin{tabular}{rr}  Turnovers & Start Pos \\ \hline  Rk (Val) & Rk (Val) \\ \hline 32 (0.09) & 32 (28) \\ 31 (0.13) & 32 (37) \\ 7 (0.21) & 31 (36) \\ 32 (0.15) & 28 (38) \\ 31 (0.12) & 32 (35) \\  ... & ... \\5 (0.21) & 1 (58) \\ 1 (0.28) & 1 (49) \\ 2 (0.26) & 4 (50) \\ 1 (0.25) & 4 (50) \\ 7 (0.21) & 3 (51) \\  \end{tabular}
 \end{tabular}
 \end{adjustbox}
    \label{tab:SeasonLongShiftsOffenseNFL}
\end{table*}

On the other hand, offenses experiencing the largest drops in points scored per drive are those complemented by the best defenses in terms of turnovers and post-turnover starting positions. During those respective years, all five defenses were ranked in the Top 4 for post-turnover starting position and no lower than 7th in their ability to generate turnovers, consistently providing their offenses with excellent scoring opportunities. It is worth noting that the negative shifts are higher in magnitude compared to the positive shifts. This disparity stems from the inherent right-skewness in the distribution of the complementary feature variable that combines the turnover indicator with the starting position. Very few defenses generate a significant number of excellent direct scoring opportunities for their offenses in a given season, while most defenses cluster near the lower end of the distribution, with the league-average value naturally being closer to that lower end. Consequently, projecting the performance of offenses complemented by elite defensive units onto a league-average unit results in a stronger downward adjustment, whereas projecting offenses complemented by poor defensive units does not produce as significant an upward adjustment. This explanation clarifies why, for example, the 2013 Kansas City Chiefs' offense experienced a much larger decrease in points per drive compared to the positive boost the 2016 Chicago Bears' offense received. The Chiefs' defense had an exceptional season in 2013 in terms of generating outstanding scoring opportunities relative to the league average, whereas the Bears' defense, although ranked the worst in the league in 2016, was not significantly worse than other defenses from a numerical perspective. When accounting for the number of drives, it is unsurprising that the 2013 Kansas City Chiefs' adjustment credited approximately $209 \times 0.13 \approx 35$ of the offense's points that season to their exceptional defense. In contrast, the 2016 Chicago Bears' offense gained about $176 \times 0.08 \approx 14$ extra points due to their ability to overcome their comparatively poor defense from complementary football standpoint.

When examining the ranking shifts, although the changes are not necessarily drastic, there are several notable adjustments. In particular, the 2012 New York Giants' offense dropping from \#3 to \#8 and the 2011 Houston Texans' offense falling from \#7 to \#13 are noteworthy. These shifts suggest that, once the quality of their complementary unit is accounted for, these offenses drop out of the Top 3 and Top 10 in a 32-team league, respectively.

In addition, Table \ref{tab:SeasonLongShiftsOffenseCFB} provides a summary of the largest positive and negative shifts in points scored per drive for the college football data. Due to the necessity of excluding one-third of the college data because of record inconsistencies, combined with the shortened 2020 season caused by the COVID-19 pandemic, we imposed a minimum games threshold to ensure a fairer comparison of shifts among teams. Without this threshold, teams that played fewer games would have a disproportionately higher chance of experiencing larger shifts. Given that a typical college football season consists of 12 games, compared to the NFL's 16 games, and accounting for the shortened 2020 season, we set the minimum at five games. This threshold ensured a respectable sample size of at least 60 drives for the vast majority of teams while avoiding the exclusion of too many teams in any given year. As a result, at least 127 teams (out of approximately 130) were included for the 2014--2019 seasons, and 86 teams were included for the shortened 2020 season.

 \begin{table*}[t]
    \centering
       \caption{The five most positive (top) and five most negative (bottom) shifts in points scored per drive by the offense over a season in the Division-I college football's Football Bowl Subdivision (FBS), along with 95\% bootstrap quantile confidence intervals, corresponding league ranking shifts, and complementary unit performance rankings that season. These shifts are relative to adjustments made only for strength of schedule and homefield factors. 
       }

       \medskip

\begin{adjustbox}{max width=\textwidth}
 \begin{tabular}{rrrcc} 
Team \ & Year \ & Drives \ & Points Scored (Per Drive) & Defense Statistics (Per Drive) \\ \hline 
 \begin{tabular}{r} \\ \\ Vanderbilt \\ Air Force \\ ARK \\ Syracuse \\ Auburn  \\ .... \\ GA South \\ Coast CAR\\ MI State \\ OH \\ VA \end{tabular} & 
 \begin{tabular}{r} \\ \\ 2019 \\ 2019 \\ 2020 \\ 2016 \\ 2015  \\ .... \\ 2015 \\ 2020 \\ 2014 \\ 2017 \\ 2018 \end{tabular} & 
 \begin{tabular}{r} \\ \\ 82 \\ 65 \\ 108 \\ 78 \\ 113  \\ .... \\ 66 \\ 68 \\ 157 \\ 105 \\ 74 \end{tabular} &
 \begin{tabular}{rr} \\ Val (Shift, 95\% CI) \ \ & Rk (Shift) \\ \hline 1.64 ($\textcolor{darkgreen}{\Uparrow}$ 0.13, [0.10, 0.17]) & 92 ($\textcolor{darkgreen}{\Uparrow}$ 8) \\ 2.91 ($\textcolor{darkgreen}{\Uparrow}$ 0.12, [0.08, 0.15]) & 12 ($\textcolor{darkgreen}{\Uparrow}$ 3) \\ 2.98 ($\textcolor{darkgreen}{\Uparrow}$ 0.12, [0.09,  0.14]) & 17 ($\textcolor{darkgreen}{\Uparrow}$ 5) \\ 1.78 ($\textcolor{darkgreen}{\Uparrow}$ 0.11, [0.09, 0.14]) & 76 ($\textcolor{darkgreen}{\Uparrow}$ 9) \\ 2.60 ($\textcolor{darkgreen}{\Uparrow}$ 0.11, [0.09,  0.13]) & 19 ($\textcolor{darkgreen}{\Uparrow}$ 6) \\ ... & ... \\2.19 ($\textcolor{red}{\Downarrow}$ 0.28, [-0.30, -0.24]) & 50 ($\textcolor{red}{\Downarrow}$ 21) \\ 2.69 ($\textcolor{red}{\Downarrow}$ 0.24, [-0.28, -0.21]) & 24 ($\textcolor{red}{\Downarrow}$ 6) \\ 2.59 ($\textcolor{red}{\Downarrow}$ 0.19, [-0.21, -0.17]) & 20 ($\textcolor{red}{\Downarrow}$ 8) \\ 2.03 ($\textcolor{red}{\Downarrow}$ 0.17, [-0.19, -0.15]) & 59 ($\textcolor{red}{\Downarrow}$ 10) \\ 1.75 ($\textcolor{red}{\Downarrow}$ 0.16, [-0.18, -0.14]) & 81 ($\textcolor{red}{\Downarrow}$ 8) \\  \end{tabular} & 
 \begin{tabular}{rr}  Turnovers & Start Pos \\ \hline  Rk (Val) & Rk (Val) \\ \hline 122 (0.13) & 131 (21) \\ 89 (0.17) & 125 (28) \\ 62 (0.20) & 123 (28) \\ 116 (0.13) & 129 (23) \\ 63 (0.19) & 127 (30) \\  ... & ... \\2 (0.33) & 5 (55) \\ 14 (0.28) & 2 (58) \\ 10 (0.26) & 2 (53) \\ 46 (0.21) & 1 (59) \\ 1 (0.32) & 19 (47) \\  \end{tabular}
 \end{tabular}
\end{adjustbox}
    \label{tab:SeasonLongShiftsOffenseCFB}
\end{table*}

It illustrates similar intuitive patterns to those in Table \ref{tab:SeasonLongShiftsOffenseNFL} for the NFL: the biggest positive shifts are experienced by offenses with the best complementary units in terms of turnover generation and post-turnover starting position, while the opposite is true for offenses undergoing the most negative shifts. Once again, it can be confirmed that while the ability to generate turnovers is important, the post-turnover starting position is even more so. A handful of offenses receiving the largest positive adjustments had defenses that were close to league average in generating turnovers but were near the bottom (no higher than 123rd out of approximately 130 teams) in terms of the starting position that followed. As for the negative adjustments, a symmetric principle applies in the case of the 2017 Ohio (OH) team: their defense ranked only 46th in turnover generation but had a stellar post-turnover starting position of 59 yards (9 yards into the opponent's territory), making life much easier for the Ohio offense. The 2018 Virginia (VA), on the other hand, illustrates how even a moderately respectable post-turnover starting position of 47 yards (ranked 19th in the league) could yield stronger complementary benefits when combined with elite turnover-generating ability, as VA's defense led the league with 0.32 turnovers per drive.

In the supplementary materials we provide the analogous analysis for the defenses, showcasing the largest adjustments in projected points allowed based on the offense's ability to avoid turnovers that put their defense in a bad starting position. The key findings symmetrically mirror those discussed for the offenses. 
For example, the 2009 Detroit Lions offense was the worst in the league in terms of both turnovers and starting position, which led to their defense receiving a decrease (which is a positive thing for a defense) of 0.19 points allowed per drive due to our projection based on a league-average complementary offense. This adjustment accounts for a total of 35 points over the 185 drives they played that season. On the other hand, the 2013 San Diego Chargers offense was one of the best at limiting turnovers (5th in the league) while also avoiding poor starting positions for their defense (2nd). As a result, our projection added 0.09 points to the points allowed per drive by their defense, indicating that they would have allowed an additional 15 points over 169 drives if complemented by a worse, league-average offense.

\begin{figure*}[t]
\centering
        \includegraphics[scale=0.8]{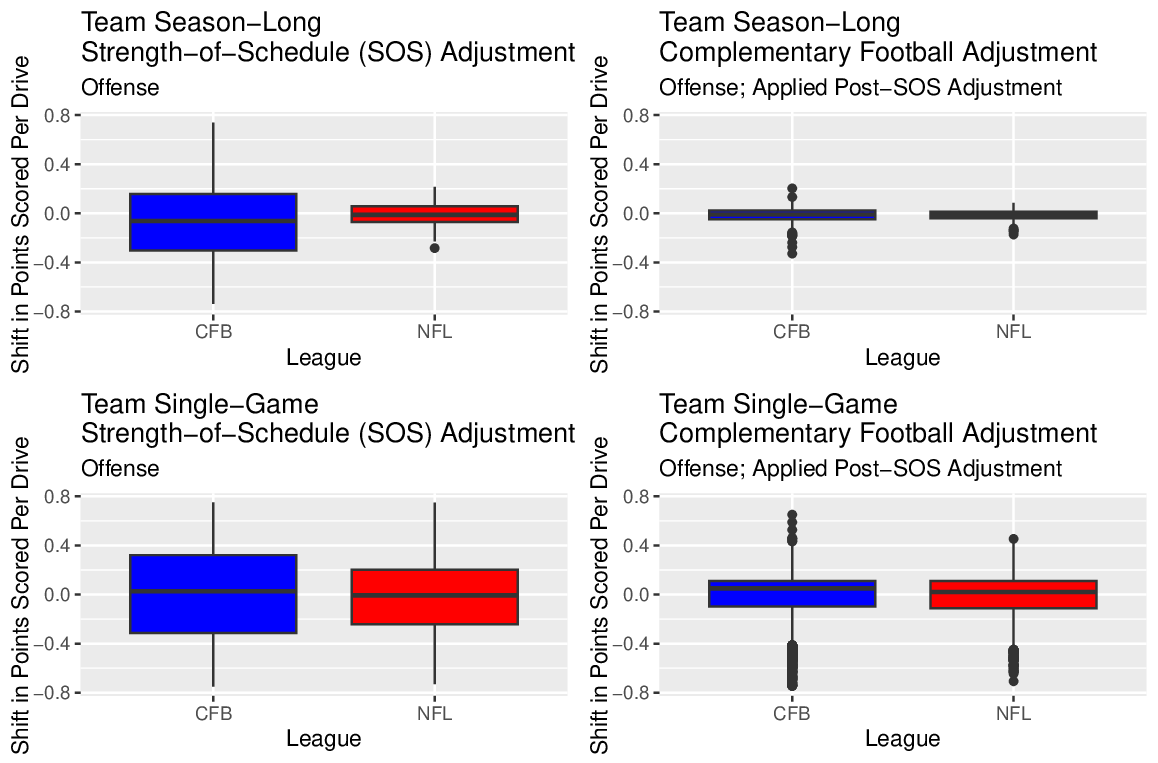}
\caption{The magnitudes of shifts in points scored per drive based on different adjustment facets (strength of schedule, complementary football) and granularities (season-long, single game).}
\label{fig:BoxplotsShifts}
\end{figure*}

Lastly, one might wonder why the projection shifts in points per drive, when adjusting for complementary football, do not appear very strong. Figure \ref{fig:BoxplotsShifts} provides some helpful context. It shows how the part of our adjustment responsible for projecting onto a common opponent (strength of schedule) drives most of the shift in values. In comparison, the complementary football projection, although still important and intuitive (as shown previously), does not cause as much movement as the strength of schedule component. This is especially clear in the season-long dynamics (top row), and more specifically for college football. The notable season-long swings when adjusting for strength of schedule in college are not surprising, given the uneven level of competition, where a handful of teams can dominate recruiting within a region. This contrasts sharply with the NFL, where the league format is designed for parity, incorporating aspects such as salary caps, draft, and other mechanisms.

Nonetheless, when we examine the single-game dynamics (bottom row of Figure 4), note that the importance of the complementary football adjustment increases, both in absolute magnitude and relative to the strength of schedule adjustment. Although it still isn’t as strong as the projection onto a common opponent, it clearly narrows the gap. 
This suggests that the complementary football adjustment is likely to be an even more useful tool when evaluating a team’s performance in a single game. The earlier example of the Kansas City Chiefs defense maintaining excellent performance in creating great scoring opportunities for their offense throughout the entire 2013 season appears to be quite rare. In contrast, playing a single excellent (or terrible) game from a complementary standpoint is a much more frequent occurrence. For tables detailing the strongest shifts in single-game points per drive scored/allowed in the NFL and college, please refer to the supplementary materials.


\subsection{Out-of-Sample Model Testing}

Lastly, we wanted to also showcase the out-of-sample predictive performance of our final selected model.

\begin{figure*}[t]
\centering
        \includegraphics[scale=0.9]{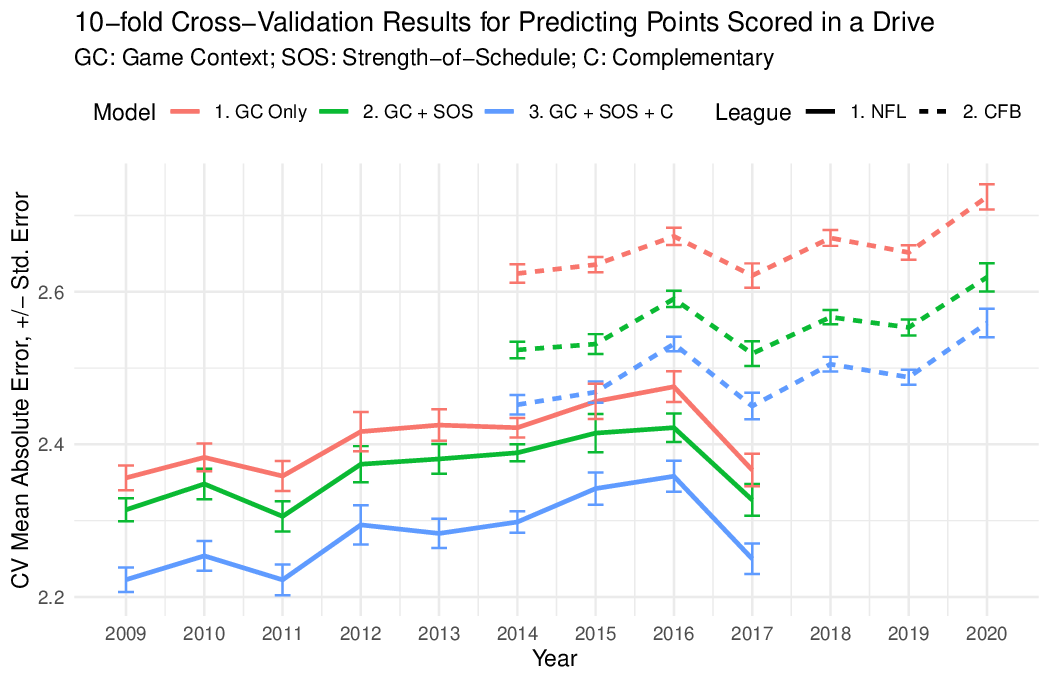}
\caption{The 10-fold cross-validation results are presented to compare ordinal regression models predicting points scored per drive with three feature sets: game context only (homefield factor, half of play, time remaining in the half, score differential), game context with strength of schedule variables (GS+SOS), and the inclusion of complementary football features, as used in our final reduced model (GS+SOS+Complem). Mean absolute error was used as the evaluation metric.}
\label{fig:10fold_CV_OutOfSample_MAE}
\end{figure*}

Figure \ref{fig:10fold_CV_OutOfSample_MAE} below depicts the results of a 10-fold cross-validation procedure conducted on both the NFL and college football data across each season under consideration. It is clear that the feature set including all three aspects (game context, strength of schedule, and complementary football) outperforms the models that exclude complementary features, confirming their importance in predicting the scoring outcome of a drive. Additionally, there is a larger gain from incorporating strength of schedule alongside game context variables for college football compared to the NFL. This further supports the intuition that team-level disparities are stronger in college football, making them more predictive of drive scoring outcomes, whereas the NFL exhibits greater parity. Finally, it is worth noting that the errors are lower for the NFL compared to college. This is most likely due to the higher inherent drive-to-drive variance in points scored in college, which makes accurate estimation more challenging. Out of curiosity, we also conducted a 10-fold cross-validation for the full feature set to compare the out-of-sample performance of our ordinal regression model with that of a classic linear regression when predicting points scored in a drive. The results, which can be found in the supplement, show a slight edge for the ordinal regression. This, combined with the greater flexibility of being able to produce probabilities for specific scoring outcomes, further supports the appropriateness of our modeling choice.

Moreover, given the ordinal nature of our model, we also assessed the calibration of our scoring category probability prediction mechanism on out-of-sample data, using a method inspired by \citep{yurko2019nflwar}. Our model was shown to be well-calibrated across all five scoring events for the vast majority of the NFL and college seasons. For more details on the procedure and the discussion of the results, see the supplementary materials. 

\section{Discussion and future work}
\label{sec:Discussion}

We have implemented a partially regularized ordinal regression approach to infer the most consistent features of complementary football that affect scoring in American football, both college and the NFL, while making sure to account for strength of schedule and home-field factor. It involved us modifying the optimization criteria and coding algorithm made available in \citep{wurmOrdinalNet} to allow for partial introduction of non-proportional odds, while constraining some factors to still have proportional odds. Without that modification, due to the need to always include all the team-level strength-of-schedule factors in our model, the computational task of estimating such a model with non-proportional odds for these factor variables would have been prohibitive. Even if the model eventually converged, it likely would have suffered from issues related to the absence of certain scoring outcomes (e.g., the rarely occurring "Safety" play) for some teams throughout the season. 

Moreover, we extended the {\em ordinalNet} software to enable the fitting of an unpenalized non-proportional odds model, which included a non-proportional odds coefficient solely for one of the categories (in our case, the "At least a field goal" category). Functionality of {\em ordinalNet} only allowed for either a full regularization grid search, which penalized all non-proportional odds coefficients and did not guarantee the inclusion of a specific subset, or an unpenalized fit that required including non-proportional odds coefficients for all response categories, rather than just a specific subset. Our extension was crucial in allowing us to fit a consistent model across all seasons for both the NFL and college football, ensuring the same adjustment mechanism for complementary football features of importance. Additionally, to conduct proper residual diagnostics for the resulting non-proportional odds models, we extended the {\em sure} package functionality that implements the surrogate residual methodology for ordinal regression models. The original methodology was only applicable to proportional odds models, where residuals across all response categories were combined into a single diagnostic, and we simply modified it to produce separate plots for the cumulative odds of each respective scoring category.

From the results relevant to the application domain, the football statistics that proved to be most important in impacting the complementary unit were non-scoring turnovers and post-turnover starting position. The nature of these effects was intuitive, indicating that a defense's turnovers and starting positions deeper in the opponent's territory have a larger positive impact on the offense's scoring production. Moreover, we leveraged these findings to create a more contextualized version of offensive and defensive team rankings, which not only projects performance onto a league-average opponent and homefield factor but also onto a league-average complementary unit. The shifts in projected values were intuitive, with offenses that had a strong complementary defense receiving a downgrade in their points scored per drive once projected onto a league-average defense, and those complemented by weaker defenses receiving an upgrade. Similarly, defenses complemented by strong offenses were downgraded, by weaker ones - upgraded. 

We also discussed how complementary football adjustments compare to strength-of-schedule adjustments, with the latter resulting in larger value shifts, especially on a season-long scale. Nonetheless, on a single-game level, the complementary football projections showed a more notable magnitude. 
As discussed earlier, it is much more difficult for a complementary unit to maintain an exceptionally great (or bad) performance throughout the entire season, while in a single game such performance is more attainable. This suggests that our adjustments can be even more impactful when evaluating a team's performance in a single game.


Lastly, we demonstrated the out-of-sample predictive improvement gained by incorporating complementary football features, alongside game context and strength-of-schedule data. This highlighted the clear improvement from accounting for complementary features, while also illustrating the differences between the NFL and college football in terms of the out-of-sample explanatory power gained from the strength-of-schedule factors.

\if0\blind
{
This work is similar to \citep{skripnikov2023partially}, but it presents a more detailed and well-thought-out modeling design, incorporating the sequential drive-by-drive nature of the game, as opposed to simply working with game totals, which makes one more vulnerable to issues such as reverse causality. Moreover, the aforementioned work did not explicitly incorporate the post-turnover starting position, hereby presenting a less comprehensive framework. Nonetheless, it is still noteworthy that the main findings are replicable compared to previous work, which also discovered turnovers as one of the most critical aspects of complementary football. To add to the discussion of the replicability, \citep{kempton2016expected} studied the expected points for rugby plays based on the starting field position of the current drive (referred to as ``possession'') and the outcome of the previous drive. Although rugby, unlike American football, has the same players involved in both offense and defense, it shares a similar sequential nature of drives and the role of starting position in the scoring outcome. \citep{kempton2016expected} found, unsurprisingly, that drives starting closer to the opposing goal line had the highest expected points scored, which strongly aligns with our results. More curiously, they also identified that drives directly following an opponent's error were more likely to lead to a score. Given that a defensive turnover can be considered an error by the offense in American football, this finding draws another parallel with our results.  However, due to the dual roles of players in rugby, it is less intuitive to adjust the offensive/defensive performance for their ``complementary unit'', further highlighting how uniquely appropriate American football is for such an application. 
} \fi

\if1\blind
{
This work is similar to [X], but it presents a more detailed and well-thought-out modeling design, incorporating the sequential drive-by-drive nature of the game, as opposed to simply working with game totals, which makes one more vulnerable to issues such as reverse causality. Moreover, the aforementioned work did not explicitly incorporate the post-turnover starting position, hereby presenting a more comprehensive framework. Nonetheless, it is still noteworthy that the main findings are replicable compared to previous work, which also discovered turnovers as one of the most critical aspects of complementary football. To add to the discussion of the replicability, \citep{kempton2016expected} studied the expected points for rugby plays based on the starting field position of the current drive (referred to as ``possession'') and the outcome of the previous drive. Although rugby, unlike American football, has the same players involved in both offense and defense, it shares a similar sequential nature of drives and the role of starting position in the scoring outcome. \citep{kempton2016expected} found, unsurprisingly, that drives starting closer to the opposing goal line had the highest expected points scored, which strongly aligns with our results. More curiously, they also identified that drives directly following an opponent's error were more likely to lead to a score. Given that a defensive turnover can be considered an error by the offense in American football, this finding draws another parallel with our results. However, due to the dual roles of players in rugby, it is less intuitive to adjust the offensive/defensive performance for their ``complementary unit'', further highlighting how uniquely appropriate American football is for such an application. 
} \fi

Lastly, we have found an especially strong effect of the identified complementary statistics on the odds of a team scoring at least a field goal, but not necessarily on the odds of scoring a touchdown. As mentioned in Section \ref{sec:FeatureSelectionStabilityAnalysis}, although a complementary defense's non-scoring turnover followed by a favorable starting field position is highly beneficial for the offense's chances to score, it typically leaves less space for the offense to operate. \citep{RedZone} analyzed data on the percentage of times offenses managed to score a touchdown when within $5$-$15$ yards of the opponent's end zone. While being closer is generally better, it is far from a guarantee, with touchdown percentages ranging from $70\%$ down to $55\%$. Scoring at least a field goal, on the other hand, is much more probable in these scenarios, given that it constitutes a relatively short kicking distance for kickers at both the college and professional levels.

For future work, we are considering an extension that would explicitly incorporate point values in the optimization procedure, rather than solely relying on the order of scoring categories. Currently, we only introduce the point values in a post-hoc fashion by weighting them with the predicted probabilities when calculating the expected points per drive, but we do not leverage them in the parameter estimation itself. Since classical regression is not appropriate in our case due to the categorical nature of our response values, we plan to develop an approach that can benefit from knowing the exact point value designations of each scoring outcome, while still accounting for the categorical underlying data-generation mechanism. Moreover, despite their relative rarity, we aim to accommodate cases where a touchdown score results in a value other than $7$, potentially introducing an additional layer of conditional response probabilities into our categorical modeling structure.

\section*{Acknowledgments}
\label{sec:Acknowledgments}

\if0\blind
{
The authors are grateful to New College of Florida for providing summer research funding. The authors used ChatGPT to edit the text for clarity, grammar and syntax, but made sure to subsequently review the text themselves and confirm the actual meaning was preserved.
} \fi

\if1\blind
{
The authors used ChatGPT to edit the text for clarity, grammar and syntax, but made sure to subsequently review the text themselves and confirm the actual meaning was preserved.
} \fi


\section*{Statements and declarations}
\label{sec:DeclarationOfInterest}

\subsection*{Declaration of interest}
We confirm that there are no known conflicts of interest associated with this publication  and there has been no financial support that could have influenced its outcome.

\if0\blind
{
\subsection*{Data and code availability}
Data and source code for this work have been made publicly available on Github via this link: \url{https://github.com/UsDAnDreS/American_Statistician_PartiallyRegularizedOrdinalReg_ComplementaryFootballPaper}.
}

\newpage
\section*{Appendix}

\subsection{Complementary Football Features Glossary}

Table \ref{tab:GlossaryCompFeatures} contains all the complementary football statistics under consideration in this work, with their respective description.

\begin{table*}[h!]
    \caption{Glossary of all considered complementary football features (measured on the previous drive). *Subtracting the punt or kickoff yardage from the return yardage. 
    }
    \centering
{\renewcommand{\arraystretch}{1.2}
    \begin{tabular}{ll}
       Name & Description \\
    \hline
    $off.ST.yards.gained$   & Yards gained by offense and special teams return \\
             $ST.return.yards.net$   & Net yards on a special teams return\footnotemark[1] \\
      $pts.scored$    & Points scored \\
       $first.downs.gained$ & Number of first downs gained \\
       $third.downs.converted$ & Number of third downs converted\\

    $turnover.nonscor$ &  Indicator of turning the ball directly over to the complementary unit, \\
    & e.g., interceptions, lost fumbles, missed field goals, turnover on downs \\

    $punt.safety$ & Indicator of a punt or a safety\\

       $n.completions$ &  Number of completed passes \\
       $n.incompletions$ & Number of incomplete passes   \\
     $n.stuffed.runs$   & Number of runs for $\leq 0$ yards \\
       $n.positive.runs$   & Number of runs for $> 0$ yards \\
      $n.negative.plays$  & Number of plays for $< 0$ yards, incl. penalties \\

    $n.sacks$ & Number of sacks \\
    $n.fumbles$ & Number of fumbles (lost and recovered)
    \end{tabular}
    }
    
    \footnotemark[1]{Subtracting the punt or kickoff yardage from the return yardage. Will be mostly negative.}
    \label{tab:GlossaryCompFeatures}
\end{table*}

\subsection{Selection Stability, With Starting Position Included}

\begin{figure*}[h]
\centering
\subfloat[Selection stability,  NFL\label{fig:NFL_SelectionStability_W_ST_POS_INTERACTION}]
{\includegraphics[scale=0.6]{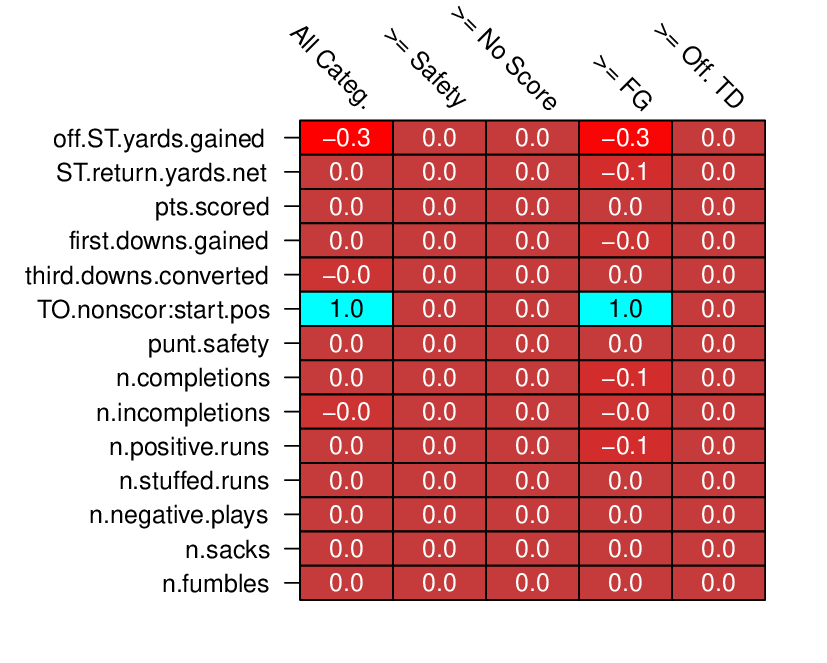}}
\subfloat[Selection stability, College\label{fig:CFB_SelectionStability_W_ST_POS_INTERACTION}]{\includegraphics[scale=0.6]{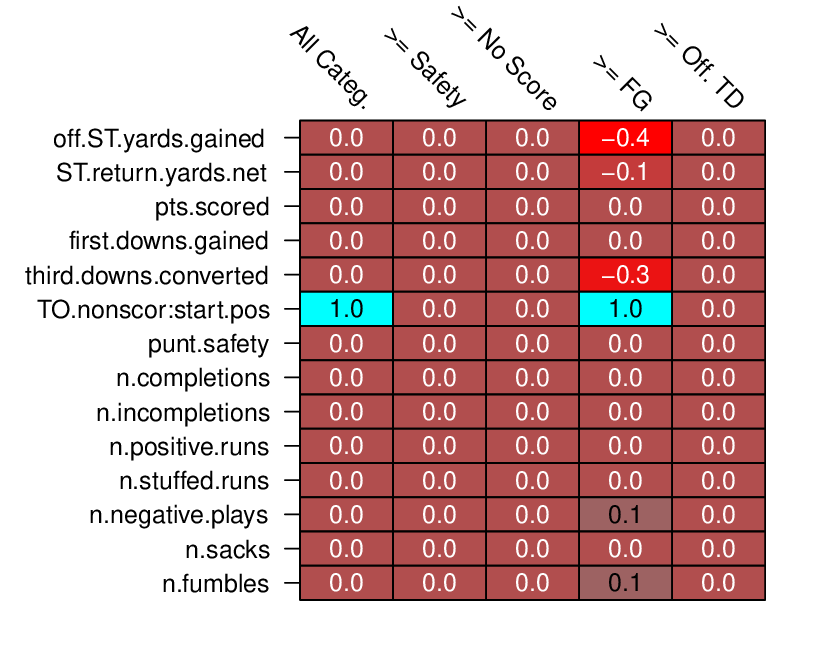}}
\caption{(a): Proportion of cross-validation replicates for 2009-2017 NFL seasons for which the respective complementary football statistic (rows) was chosen in how it impacts scoring on subsequent drive, including the coefficient sign. This feature set includes the term combining starting position and turnover indicator ({\em {TO.nonscor:start.pos}}). Each statistic has a proportional (first column) and non-proportional (columns 2-5, by response category) odds coefficient. (b) Same, but for the 2014-2020 FBS college football seasons.
}
\label{fig:fig_SelectionStability_W_ST_POS_INTERACTION}
\end{figure*}

\bibliographystyle{chicago}
\bibliography{bibliography}

\end{document}